# Humans: the hyper-dense species

**Joseph R Burger[1], Vanessa P Weinberger[2,4], Pablo A Marquet[2,3,4,5]**
Correspondent email: jrburger@email.unc.edu

[1] Department of Biology, University of North Carolina, Chapel Hill, USA

[2] Departamento de Ecología, Facultad de Ciencias Biológicas, Pontificia Universidad Católica de Chile, Alameda 340, Santiago, Chile

[3] The Santa Fe Institute, 1399 Hyde Park Road, Santa Fe, NM 87501, USA

[4] Instituto de Ecología y Biodiversidad (IEB), Santiago Chile

[5] Laboratorio Internacional de Cambio Global (LINCGlobal) & Centro de Cambio Global UC, Pontificia Universidad Católica de Chile, Alameda 340, Santiago, Chile

**Humans, like all organisms, are subject to fundamental biophysical laws. Van Valen predicted that, because of zero-sum dynamics, all populations of all species in a given environment flux the same amount of energy on average[1]. Damuth's 'energetic equivalence rule' supported Van Valen´s conjecture by showing a tradeoff between few big animals per area with high individual metabolic rates compared to abundant small species with low energy requirements[2]. We use established metabolic scaling theory to compare variation in densities and individual energy use in human societies to other land mammals. We show that hunter-gatherers occurred at lower densities than a mammal of our size. Most modern humans, in contrast, concentrate in large cities at densities that are up to four orders of magnitude greater than hunter-gatherers yet cities consume up to two orders of magnitude greater energy per capita. Today, cities across the globe flux greater energy than net primary productivity on a per area basis. This is possible through enormous fluxes of energy and materials across urban boundaries to sustain hyper-dense, modern humans. The metabolic rift with nature created by hyper-dense cities supported by fossil fuel energy poses formidable challenges for establishing a sustainable relationship on a rapidly urbanizing, yet finite planet.**

All populations, including humans, are sustained by fluxes of energy and materials from a finite environment. Physical constraints on biological design result in ubiquitous and predictable allometric scaling laws[3]. These take power law form where some trait of interest ($R$), scales with body size ($M$),

$$R \propto cM^{\beta} \quad \text{(eqn 1)}$$

where $\beta$ is the exponent and $c$ is the intercept. Metabolic scaling theory predicts quarter-power exponents for rates and quantities across many levels of biological organization[3,4] including whole organism and mass-specific metabolic rates, which scale as $\beta \approx ¾$ and $\beta \approx -¼$, respectively. Allometric parameters (i.e., intercept and slope) can be predicted theoretically and evaluated empirically to form a quantitative framework to carry out meaningful comparisons across scales from cells and organisms[5] to human societies[6,7]. Using this framework as a reference we aimed to understand unique aspects of human ecology and to quantify the extent to which the human species has departed from the energetic constraints that keep all other species in check.

An important ecological implication of metabolic scaling is the inverse relationship between body size and density. Because individual metabolic rate ($E_i$), scales predictably with size[4]

$$E_i \propto M^{-3/4} \quad \text{(eqn 2)},$$

the maximum number of individuals per unit area ($D_{max}$), scales as the inverse of individual energy demands[2,8]

$$D_{max} \propto M^{-3/4} \quad \text{(eqn 3)}.$$

The result is a tradeoff between size and abundance. Abundance *per se* is not limited by body mass but instead by the energy required to support an individual, $E_i$, of a given body size. Rearranging the allometric relationships in Eqn 2 & 3, theory predicts that

$$D_{max} \propto E_i^{-1} \quad \text{(eqn 4)}$$

and population energy flux, $E_p$, calculated as the product of $E_i \cdot D_{max}$, is invariant across species,

$$E_p \propto E_i^0 \text{ (eqn 5)}$$

with $\beta = 0$. This 'energetic-equivalence rule' (EER[8]) links individual metabolic requirements to population energy use in space and time. This is consistent with the existence of a zero sum game for energy use as predicted by Van Valen[1] and quantified in local mammal communities[9]. Unique to industrial humans, however, is the capacity to harness extra-metabolic energy in the form of renewables and fossil fuels to power modern agricultural-technological-industrial lifestyles. Among human societies, individual energy consumption varies from ~120 watts of biological metabolism — the equivalent of ~2500 kilocalories per day — in hunter-gatherers to more than 10,000 watts in the most developed nations[7,10]. So clearly, humans have deviated from other species in their energy use.

**Results and Discussion**

Figure 1 shows that herbivorous land mammals support theoretical predictions where density decreases proportionally with individual energy requirements (slope = -1.08; 95% CI: -0.88, -1.27). Hunter-gatherers, in contrast, occur at densities lower than expected based on other land mammals (ANOVA, F-interaction=6.37, p <0.001, Tukey post-hoc test). The trophic position where an organism feeds in the food web explains additional variation in the densities of land mammals and hunter-gatherers where densities decrease with higher trophic levels (Supplemental materials; Table S1).

Across cities, density scales negatively with increasing per capita energy requirements consistent with theoretical predictions and similar to the scaling of land mammals. However, modern city dwellers occur at densities that are four orders of magnitude greater than hunter-gatherers and other land mammals (Figure 1; Table S1) even though they consume one to two-orders of magnitude greater energy, per capita. The highest density city in our data (Dhaka, Bangladesh with 44,000 ind/km$^2$) now surpasses the highest density wild rodent (Townsend's vole with 34,349 ind/km$^2$). The slope for urban humans (-0.44; 95% CI: -0.32, -0.55) is shallower than the theoretical expectation (slope of -1) and estimated empirical slope from land mammals (Table S1).

Consistent with Van Valen's zero-sum prediction and the EER, the energy flux by herbivorous land mammal populations ($10^{-1}$ to $10^4$ watts/km$^2$) is invariant with individual energy use (Fig 2) with a slope indistinguishable from 0 (-0.08, 95% CIs: -0.28, 0.12). Energy flux by hunter-gatherers range from $10^2$ to $10^3$ watts/km$^2$ and is lower than other land mammals (ANOVA, F-value interaction=10.01, p <0.001, Tukey post-hoc test). Members of pre-industrial societies fluxed greater energy ($10^4$ to $10^5$ watts/km$^2$) than hunter-gatherers and other land mammals on average but less than cities. Energy flux in modern cities ranges from $10^5$ to $10^8$ watts/km$^2$ and surpasses global terrestrial primary productivity ($10^5$ watts/km$^2$ global avg). Unique to urban humans is the positive relationship between population energy flux and per capita energy requirements (slope=0.56 [95% CI: 0.44, 0.68]), whereas other land mammals show a slope indistinguishable from 0 as theory predicts (Figure 2). See supplemental for additional analyses by trophic levels.

Just like all mammals, hunter-gatherers are faced with the challenge of meeting metabolic demands from the local environment to power their lifestyles and sustain their populations[11]. Pre-industrial societies lived at greater densities than hunter-gatherers through the use of agriculture. Humans in modern cities live at densities much greater than those in hunter-gatherers and pre-industrial societies and consume up to two orders of magnitude greater energy per capita than caloric needs alone. This shows that the rapid rise in human densities, which has occurred in less than 10,000 years, is coincident with innovations in food production and extra-metabolic energy use from renewable and fossil fuels[12].

One salient characteristic of complex human social systems is our ability to copy the behavior of others to propel a cumulative cultural evolution (CCE[13]). It is easy to see that this process speeds up with greater population density and information flow as a result of greater energy throughput. The positive slope for population energy flux with per capita energy use in Figure 2 is a consequence of a shallower slope in Figure 1. This suggests that increased per capita energy use – which is tightly coupled to economic growth[7] – results in economies of scale by packing more individuals in a given area. Population density—through its effect on CCE—is a major driver of innovation[6,14] and increased social complexity[11]. These processes can generate positive feedbacks and a runaway process of cultural niche construction[15] contributing to the rapid divergence of humans from other species and the rise in human densities from hunter-gatherers to agriculturalists to modern cities. This unique aspect of human ecology is a result of the Malthusian-Darwinian dynamic that drives species to maximize power when innovations allow[16].

Throughout this process human societies have become increasingly decoupled from local environmental constraints and uncertainties. Technologies that increase resource production and global trade networks that offset imbalances in resource supply and demand provide an enormous buffer from local environmental constraints and perturbations such as drought and other human and natural disasters[17]. However, hyper-dense cities are only possible due to the fluxes of vast quantities of energy, materials, and information across city boundaries to offset resource sinks and maintain dense, urban lifestyles[18]. The energy flux required to sustain hyper-dense cities now surpasses background levels of net primary productivity on a per area basis (Figure 2) and is largely (~85% globally) in the form of carbon-based fossil fuels[7]. Increasing scarcity of essential resources including fossil fuels[7], water[20], and nutrients such as phosphorous[18] pose formidable challenges for continued urbanization and high-density cities. The cumulative impact of hyper-dense cities may surpass planetary tipping points[21] having rippling effects at multiple scales.

Our human macroecological approach offers illuminating insights into how humans have rapidly diverged from other species in the course of our unique CCE. Through the use of extra-metabolic energy, modern humans have escaped the energetic constraints that are imposed on all other species. Our approach also highlights the extraordinary densities that humans have obtained through the continued capacity to harness energy reserves from the planet in the form of sunlight[22] to energy stored on geological time scales[19]. Whether density-dependent innovations will continue to outpace resource constraints on human population is uncertain. What is certain is that the steep metabolic rift with nature[23,24,*sensu* 25] created by the massive energy subsidies required to support growing, hyper-dense cities poses formidable challenges to achieving sustainability in a post-fossil fuel world.

**METHODS**
*Data.* We use global data from ecological, archeological, demographic, and economic sources (Supplemental data). Data for metabolic rates and densities for 249 land mammals are from the PANTHERIA species-level database[26]. Densities for 339 hunter-gatherers are available in Binford's comparative ethnographic and environmental database[27]. Hunter-gatherer lifestyles are powered by biological metabolism estimated at 120 watts (~2500 kilocalories/day following[7,10]). Densities for pre-industrial societies (n=4) are from[28] and energy use estimated at 600 watts[10].

The EER applies to primary consumers[2,8], so we plot herbivorous land mammals (n=74) and hunter-gatherers that obtain greater than two-thirds of their diet from plants (i.e., gatherers, N=31) in figures 1 and 2. It is well known that species at higher trophic levels occur at lower densities (e.g., [29]) and hunter-gatherers are no exception with gatherers occurring at higher densities than omnivores, which occur at higher densities than carnivores (See supplemental materials for additional trophic analyses).

The EER predicts maximum animal densities[2]. So, we use the densest city per country for which data are available from Demographia World Urban Areas (demographia.com; website has updated lists). The data consists of census conducted between 2000 and 2014. In contrast to hunter-gatherers, urban human lifestyles are powered by both biological metabolism and extra-metabolic energy in the form of renewables and fossil fuels[7,10]. However, a global database on metabolic (i.e., caloric) and extra-metabolic energy use for cities is not available. So we estimated per capita energy use by combining country-level data on food consumption per country (kcal per capita per day) from the Food and Agriculture Organization from of the United Nations (FAO) (http://faostat3.fao.org/) with extra-metabolic energy from The World Bank Indicators (http://data.worldbank.org/indicator; website has updated data). We use only one city (the densest) per country in our analyses although more cities are listed in the demographia dataset. This assumes greater variation among countries than within, which is supported by studies showing that resource use and waste production scale linearly (e.g., constant per capita) with city size within countries[6,30].

*Statistical Analyses.* We determined the relationship between population density and per capita energy requirements separately for land mammals and urban humans using ordinary least squares regression of log transformed variables. We compared 95% confidence intervals of allometric parameters from linear models of the log-log relationships for land mammals and modern cities in order to evaluate theoretical expectations (slope = -1).

We conducted Analysis of Covariance (ANCOVA), with per capita energy use as the covariate, density as the dependent variable, and trophic level or human versus land mammal as fixed factors. Comparisons were made between: (i) land-mammals by trophic levels (ANCOVA), (ii) land-mammals and hunter-gatherers considering trophic level as a second factor (ANOVA), and (iii) land mammals (with trophic level as a cofactor) versus urban humans (ANCOVA).

We calculated population energy requirements ($E_p$) as the product of population density and per capita energy use to test the EER that all populations flux the same amount of energy per unit area as a consequence of the zero-sum. We conducted ANCOVAs and ANOVAS similar to above with $E_p$ as the dependent variable.

**Acknowledgements**: We thank Fernando Alfaro, Bill Burnside, Jim Brown, Melanie Moses, Allen Hurlbert and members of the Human Macroecology Group at UNM for helpful suggestions and discussion. An initial version of figure 1 was stimulated during the Santa Fe Institute and Universidad de Desarrollo Complex Systems Summer School in Zapallar, Chile. JRB is supported by a Carolina Postdoctoral Fellowship for Faculty Diversity. VPW is supported by Programa Capital Humano Avanzado from CONICYT and Proyectos IEB ICM- MINECON, P05-002 and PFB-23.

**Contributions**: All authors designed the study. JRB and VPW compiled and analyzed the data. All authors contributed to writing the manuscript.

**Data**: available by email request (Robbie Burger: jrburger@email.unc.edu)

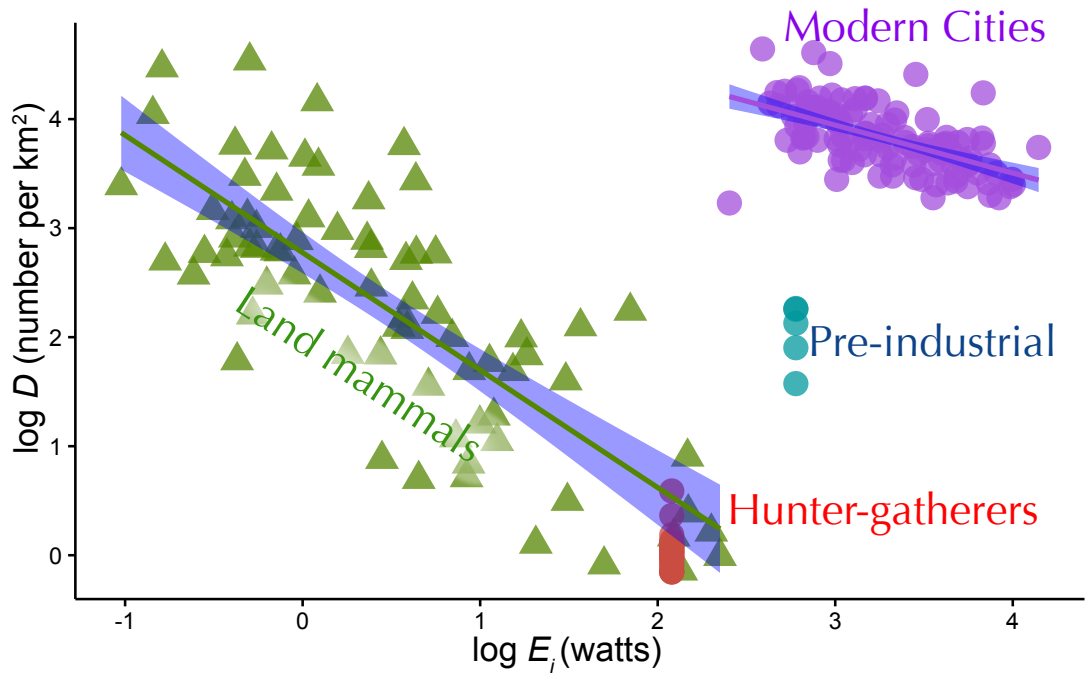

**Figure 1. The log of population density (*D*) as a function of log individual energy use (*$E_i$*) for human populatons (circles) and other land mammals (triangles).** Red circles represent vegetarian hunter-gatherers (n=31), blue circles are pre-industrial societies (n=4), purple circles are modern cities (n=163), and green triangles are other herbivorous land mammal species (n=74). Energy use for hunter-gatherers is estimated at 120 watts ~ 2500 kcals/day following[10]. Energy use for pre-industrial societies is estimated at ~600 watts from[10]. Note that the slope for cities is shallower than herbivorous land mammals, which support theoretical predictions of -1. See supplemental materials for additional details and data sources.

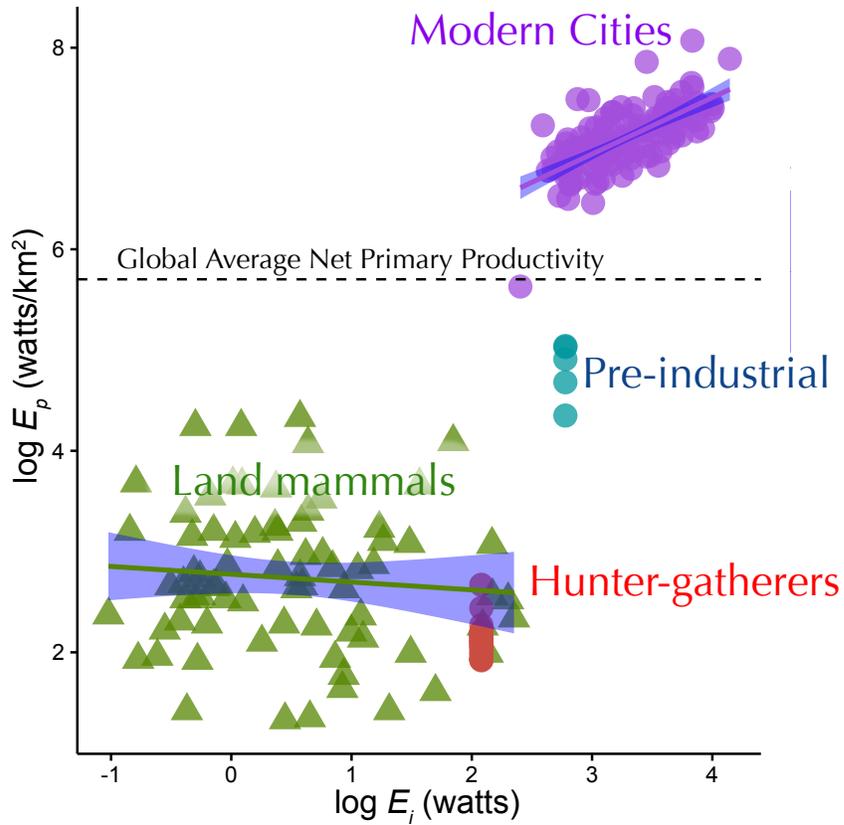

**Figure 2. Population energy flux ($E_p$) as a function of individual energy use ($E_i$) for human populations (circles) and other land mammals (triangles).** Red circles represent vegetarian hunter-gatherers (n=31), blue circles are pre-industrial societies (n=4), purple circles are modern cities (n=163), and green triangles are other herbivorous land mammal species (n=74). $E_p$ is estimated as the product of density (individuals/km$^2$) and $E_i$ (watts). Note that population energy use for herbivorous land mammals does not vary with individual energy use supporting theoretical expectations (slope = 0), whereas urban cities increase (positive slope). The dashed line represents the terrestrial average net primary productivity for the planet from[22].

Supplementary Materials for

# Humans: the hyper-dense species

**Joseph R Burger[1], Vanessa P Weinberger[2,4], Pablo A Marquet[2,3,4,5]**
Correspondent email: jrburger@email.unc.edu

[1] Department of Biology, University of North Carolina, Chapel Hill, USA

[2] Departamento de Ecología, Facultad de Ciencias Biológicas, Pontificia Universidad Católica de Chile, Alameda 340, Santiago, Chile

[3] The Santa Fe Institute, 1399 Hyde Park Road, Santa Fe, NM 87501, USA

[4] Instituto de Ecología y Biodiversidad (IEB), Santiago Chile

[5] Laboratorio Internacional de Cambio Global (LINCGlobal) & Centro de Cambio Global UC, Pontificia Universidad Católica de Chile, Alameda 340, Santiago, Chile

**Supplemental Materials include:**
Trophic analyses
Additional results
Figures S1 and S2
Tables S1 and S2
References
Data table

**Trophic level analyses**

Trophic energetics are known to influence abundance in addition to body size (e.g.,[1–4]). So we conducted a separate analysis comparing density and energy use in land mammals and hunter-gatherers by trophic levels (Fig S1 and S2). Discrete trophic levels for land mammals are from the PANTHERIA species-level database[5]. Percent diet from hunting, fishing, or gathering are available for hunter-gatherers in Binford[6]. We classified meat-eaters with ≥ 66% of diet from hunting and fishing as carnivores (n=163), mixed-diets with ≥ 33% but < 66% from plants as omnivores (n=145), and ≥ 66% of diet from plants as herbivores (n=31).

We determined the relationship between population density and per capita energy requirements separately for land mammals and urban humans using ordinary least squares regression of log transformed variables similar to [7–9]. We compared 95% confidence intervals of the slopes from linear models of the log-log relationships for land mammals and modern cities in order to evaluate theoretical expectations (slope = -1) from empirical observations.

To test for differences in density among trophic levels for land mammals, we conducted Analysis of Covariance (ANCOVA), with per capita energy use as the covariate and trophic level as a fixed factor. To compare land mammals and hunter-gatherers, we also conducted Analysis of Variances (ANOVA) with trophic level as a second factor. To compare modern cities with other land mammals we conducted a second ANCOVA similar to the above analysis using only land mammals.

To test the null hypothesis that all populations flux the same amount of energy per unit area as a consequence of the zero-sum, we calculated population energy requirements as the product of population density and per capita energy use. We conducted ANOVAs and ANCOVAs similar to above analyses but with population energy flux ($E_p$) as the response variable.

**Trophic level results**

*Density and energy use by trophic levels*

Density decreases with higher trophic levels in land mammals and hunter-gatherers are no exception. The slopes for each trophic level for land mammals (Fig S1) are significantly different (Table S1, ANOVA, F-value interaction= 6.49, p-value < 0.001, Tukey post-hoc test). The slope for herbivores is statistically indistinguishable from theoretical expectations of -1 (-1.08 [CI: -0.879, -1.27]) (Table S1) and higher trophic levels show significantly steeper slopes than theory predicts: -1.29 for omnivores and -1.63 for carnivores (Table S1). This is likely due to the greater uncertainty in food availability in space and time for organisms at higher trophic levels and has been discussed elsewhere[3,4]. Shifts in the elevation (intercepts) of slopes for herbivores, omnivores, and carnivores is roughly an order of magnitude and corresponds to an approximate 10% energy transfer between trophic levels[1].

A two way ANOVA comparing land mammals and hunter-gatherers by trophic level shows a significant interaction between factors (ANOVA, F-value interaction= 6.37, p-value < 0.001), with each hunter-gatherer, by trophic level acquiring densities significantly lower than their land mammal counterparts (Tukey post-hoc test, p-values < 0.001 for all level comparisons). Hunter-gatherers also show significant trophic effects (ANOVA, p-value < 0.001 for all group comparisons, Tukey post-hoc test), with predominately vegetarians achieving the highest densities, followed by omnivores and carnivores (Figure S1). It is important to note that none of the 339 hunter-gatherers reported in Binford are exclusively vegetarian which may explain why they occur at lower densities than their land mammal counterparts in all tests. Modern cities are also different from land-mammals when compared to trophic categories (ANCOVA, F-value interaction= 16.06, p-values < 0.001, Tukey post-hoc test) with cities having shallower slopes (Table S1).

*Energy flux and trophic levels*

Comparing energy flux per unit area among land mammals by trophic level we find that herbivorous land mammals have a slope indistinguishable from 0 (-0.08 [CI: -0.28,0.12]) as expected by theory and flux greater energy than predominately vegetarian hunter-gatherer populations (ANOVA, F-value interaction=6.37, p-values<0.001, Tukey post-hoc test; Table S2). For land mammals, higher trophic levels show significantly steeper slopes (Table S2; ANCOVA, F-value interaction= 6.49, p-value < 0.001, Tukey post hoc test) and population energy flux in omnivorous and carnivorous hunter-gatherers are indistinguishable from their land mammal counterparts (ANOVA, F-value interaction= 10.01, p-values<0.001, Tukey post-hoc test). When comparing land mammals to modern cities, we find that modern cities are significantly different from all trophic levels (ANCOVA, F-value interaction= 16.05, p-value<0.001, Tukey post-hoc test). This is evident by the unique positive slope compared to other land-mammals in Figure 2.

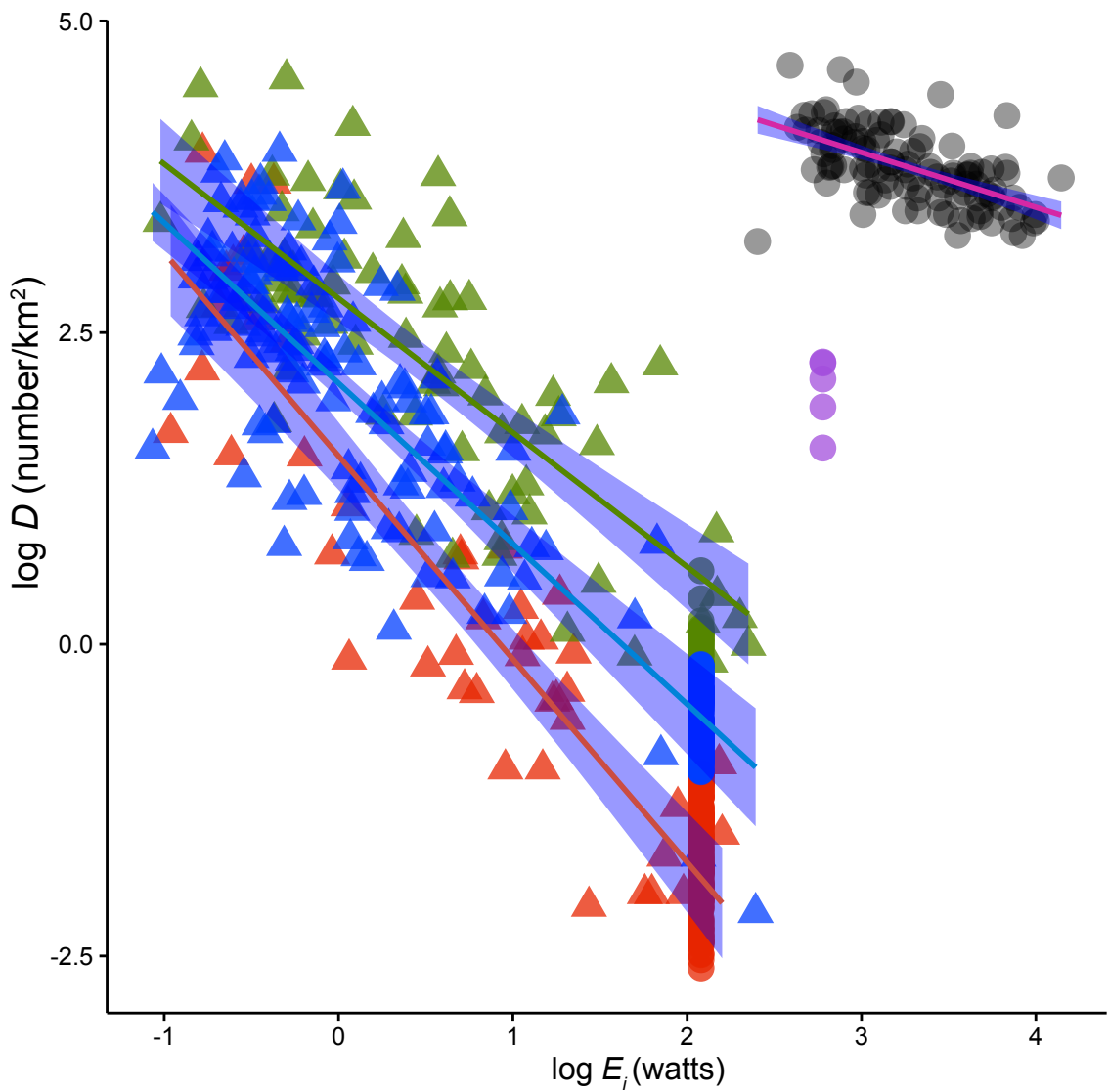

**Figure S1.** Log density versus log individual energy use for humans (circles) and other land mammals (triangles) distinguished by trophic level: red = carnivores, blue = omnivore, green = herbivores, purple = pre-industrial agriculturalists, and grey = city dweller.

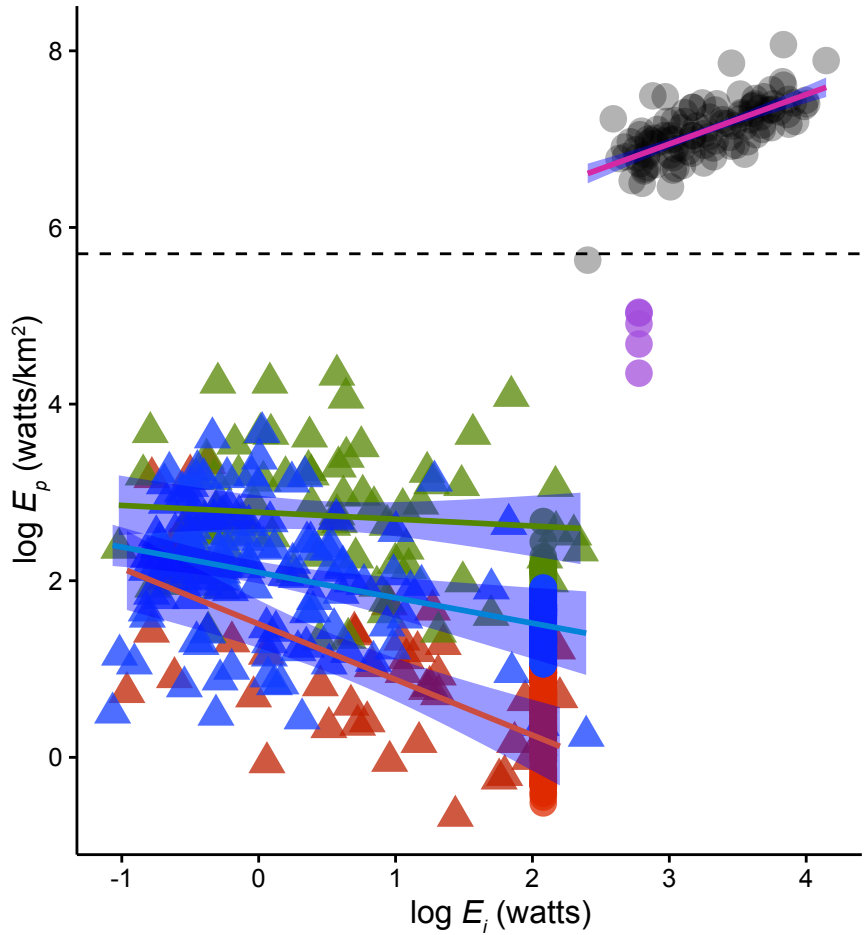

**Figure S2.** Population energy flux (the product of individual energy use and density) versus individual energy use for human societies (circles) and other and land mammals (triangles) distinguished by trophic level: red = carnivores, blue = omnivores, green = herbivores, purple = pre-industrial agriculturalists, and grey = city dweller. The dashed line represents the terrestrial average net primary productivity for the planet from[22].

**Table S1: Scaling parameters with 95% confidence intervals by category for Fig S1. Herbivore slope is indistinguishable from theoretical expectations of -1.**

| Category | R-adjusted | P-value | Slope | Intercept | Degrees of freedom |
|---|---|---|---|---|---|
| Cities | 0.33 | 2.04e-11 | -0.44 [-0.32, -0.56] | 5.26 [4.88, 5.42] | 109 |
| Herbivores | 0.62 | 2.2e-16 | -1.08 [-0.879, -1.27] | 2.77 [2.58, 2.96] | 73 |
| Omnivores | 0.59 | 2.2e-16 | -1.29 [-1.1, -1.48] | 2.09 [1.97, 2.21] | 127 |
| Carnivores | 0.80 | 2.2e-16 | -1.63 [-1.39, -1.87] | 1.51 [1.25, 1.76] | 43 |
| All land mammals | 0.57 | 2.2e-16 | -1.30 [-1.16, -1.44] | 2.19 [2.07, 2.31] | 247 |

**Table S2: Slopes and intercepts with (95% confidence intervals) for Plot 2. Zero-sum theory predicts a slope of zero and is supported by herbivores.**

| Category | R-adjusted | P-value | slope | intercept | Degrees of freedom |
|---|---|---|---|---|---|
| Cities | 0.45 | 5.47e-16 | 0.56 [0.44, 0.68] | 5.27 [4.89, 5.68] | 109 |
| Herbivores | -0.01 | 0.439 | -0.08 [-0.28, 0.12] | 2.78 [2.60, 2.84] | 73 |
| Omnivores | 0.01 | 0.0031 | -0.29 [-0.49, -0.02] | 2.09 [1.97, 2.21] | 127 |
| Carnivores | 0.37 | 5.05e-06 | -0.63 [-0.87, -0.39] | 1.51 [1.25, 1.76] | 43 |
| All land mammals | 0.06 | 2.95e-05 | -0.30 [-0.44, -0.16] | 2.19 [2.07, 2.31] | 247 |